\documentclass[conference,10pt]{IEEEtran}
\IEEEoverridecommandlockouts
\usepackage{amsmath,amssymb,amsfonts,mathtools}
\usepackage{bm}
\usepackage{siunitx}
\usepackage{array,booktabs,multirow,tabularx}
\usepackage[table,svgnames,dvipsnames]{xcolor}
\newcolumntype{P}[1]{>{\centering\arraybackslash}m{#1}}
\newcolumntype{L}[1]{>{\raggedright\arraybackslash}m{#1}}
\newcolumntype{Q}[1]{>{\columncolor{Gray}\centering\arraybackslash}m{#1}}
\usepackage[flushleft]{threeparttablex}

\usepackage{graphicx}
\usepackage{float}
\usepackage[export]{adjustbox}
\usepackage[font=small,labelfont=bf,justification=centering]{caption}
\usepackage{subcaption} 
\usepackage{algorithm}
\usepackage{algorithmic}

\usepackage{textcomp}
\usepackage{url}
\usepackage{stfloats}
\usepackage{verbatim}
\usepackage{enumitem}
\usepackage{multicol}
\usepackage{epstopdf}

\usepackage{cite}

\usepackage{glossaries}
\makeglossaries
\newacronym{ai}{AI}{Artificial Intelligence}
\newacronym{aoa}{AOA}{angle-of-arrival}
\newacronym{aod}{AOD}{angle-of-departure}
\newacronym{ap}{AP}{access point}
\newacronym{asic}{ASIC}{application specific integrated circuit}
\newacronym{awgn}{AWGN}{additive white Gaussian noise}
\newacronym{bs}{BS}{base station}

\newacronym{ce}{CE}{channel estimation}
\newacronym{cw}{CW}{continuous wave}
\newacronym{cdf}{CDF}{cumulative distribution function}
\newacronym{ced}{CED}{cumulative energy density}
\newacronym{clk}{CLK}{clock}
\newacronym{csi}{CSI}{channel state information}
\newacronym{csp}{CSP}{contact service point}
\newacronym{cpu}{CPU}{central processing unit}
\newacronym{crlb}{CRLB}{Cram\'er-Rao lower bound}
\newacronym{dlt}{DLT}{Distributed Ledger Technology}
\newacronym{dp}{DP}{differencial privacy}
\newacronym{ram}{RAM}{random-access memory}
\newacronym{ecc}{ECC}{elliptic curve cryptography}
\newacronym{ecdlp}{ECDLP}{elliptic curve discrete logarithm problem}
\newacronym{ecsp}{ECSP}{edge computing service point}
\newacronym{eirp}{EIRP}{equivalent isotropically radiated power}
\newacronym{em}{EM}{electromagnetic}
\newacronym{en}{EN}{energy neutral}
\newacronym{e2e}{E2E}{End-to-End}
\newacronym{pe}{PE}{processing engine}
\newacronym{rtl}{RTL}{register transfer level}
\newacronym{hls}{HLS}{high-level synthesis}
\newacronym{fa}{FA}{federation anchor}
\newacronym{fhss}{FHSS}{frequency hopping spread spectrum}
\newacronym{fl}{FL}{federated learning}
\newacronym{fpga}{FPGA}{Field Programmable Gate Array}
\newacronym{gdpr}{GDPR}{general data protection regulation}
\newacronym{ic}{IC}{integrated circuit}
\newacronym{ioe}{IoE}{Internet of Everything}
\newacronym{iot}{IoT}{Internet of Things}
\newacronym{id}{ID}{information decoding}
\newacronym{kpi}{KPI}{key performance indicator}
\newacronym{lca}{LCA}{life cycle assessment}
\newacronym{ls}{LS}{least squares}
\newacronym{lis}{LIS}{large intelligent surface}
\newacronym{liion}{Li-ion}{lithium-ion}
\newacronym{los}{LoS}{line-of-sight}
\newacronym{nlos}{NLoS}{non-line-of-sight}
\newacronym{lmo}{LMO}{lithium ion manganese oxide}
\newacronym{mf}{MF}{matched filter}
\newacronym{ml}{ML}{machine learning}
\newacronym{mse}{MSE}{mean square error}
\newacronym{mmwave}{mmWave}{millimeter wave}
\newacronym{mimo}{MIMO}{multiple-input multiple-output}
\newacronym{dmimo}{D-MIMO}{distributed MIMO}
\newacronym{miso}{MISO}{multiple-input single-output}
\newacronym{mpc}{MPC}{multipath component}
\newacronym{mrc}{MRC}{maximum ratio combining}
\newacronym{mrt}{MRT}{maximum ratio transmission}
\newacronym{ofdm}{OFDM}{orthogonal frequency-division multiplexing}
\newacronym{ota}{OTA}{over-the-air}
\newacronym{ii}{II}{Initiation interval}
\newacronym{nb}{NB}{narrowband}
\newacronym{nr}{NR}{New Radio}
\newacronym{dfg}{DFG}{Data Flow Graph}
\newacronym{pa}{PA}{power amplifier}
\newacronym{pdp}{PDP}{power delay profile}
\newacronym{per}{PER}{packet error rate}
\newacronym{pet}{PET}{privacy enhancing technology}
\newacronym{pki}{PKI}{public key infrastucture}
\newacronym{pc}{PC}{pilot count}
\newacronym{qrrls}{QR-RLS}{QR decomposition based recursive least squares}
\newacronym{re}{RE}{radio element}
\newacronym{rf}{RF}{radio frequency}
\newacronym{rfid}{RFID}{radio frequency identification}
\newacronym{ris}{RIS}{reflective intelligent surface}
\newacronym{rms}{RMS}{root mean square}
\newacronym{rls}{RLS}{recursive least squares}
\newacronym{rss}{RSS}{received signal strength}
\newacronym{rssi}{RSSI}{received signal strength indicator}
\newacronym{rw}{RW}{RadioWeaves}
\newacronym{sa}{SA}{synchronization anchor}
\newacronym{sdg}{SDG}{Sustainable Development Goal}
\newacronym{sdn}{SDN}{software-defined network}
\newacronym{se}{SE}{spectral efficiency}
\newacronym{sinr}{SINR}{signal-to-interference-plus-noise ratio}
\newacronym{siso}{SISO}{single-input single-output}
\newacronym{slam}{SLAM}{simultaneous localization and mapping}
\newacronym{snr}{SNR}{signal-to-noise ratio}
\newacronym{srls}{SRLS}{standard recursive least squares}
\newacronym{svd}{SVD}{singular value decomposition}
\newacronym{evd}{EVD}{Eigenvalue decomposition}
\newacronym{sp}{SP}{service point}
\newacronym{sram}{SRAM}{static random-access memory}
\newacronym{tdd}{TDD}{Time division duplexing}
\newacronym{tdoa}{TDOA}{time-difference-of-arrival}
\newacronym{toa}{TOA}{time-of-arrival}
\newacronym{ue}{UE}{user equipment}
\newacronym{uhf}{UHF}{ultra high frequency}
\newacronym{ula}{ULA}{uniform linear array}
\newacronym{ulp}{ULP}{uplink pilot}
\newacronym{uld}{ULD}{uplink data}
\newacronym{upa}{UPA}{uniform planar array}
\newacronym{ura}{URA}{uniform rectangular array}
\newacronym{uwb}{UWB}{ultrawideband}
\newacronym{wb}{WB}{wideband}
\newacronym{wpt}{WPT}{wireless power transfer}
\newacronym{zf}{ZF}{zero forcing}
\newacronym{lpt}{LPT}{laser power transfer}
\newacronym{rfpt}{RFPT}{radio frequency power transfer}
\newacronym{ipt}{IPT}{inductive power transfer}
\newacronym{cpt}{CPT}{capacitive power transfer}
\newacronym{apt}{APT}{acoustic power transfer}
\newacronym{cfo}{CFO}{carrier frequency offset}
\newacronym{lo}{LO}{local oscillator}
\newacronym{if}{IF}{intermediate frequency}
\newacronym{duc}{DUC}{digital up conversion}
\newacronym{ddc}{DDC}{digital down conversion}
\newacronym{dsp}{DSP}{digital signal processing}
\newacronym{vco}{VCO}{voltage-controlled oscillator}
\newacronym{pll}{PLL}{phase-locked loop}
\newacronym{rt}{RT}{ray-tracing}
\newacronym{lna}{LNA}{low noise amplifier}
\newacronym{dac}{DAC}{digital-to-analog converter}
\newacronym{lut}{LUT}{Lookup Table}
\newacronym{adc}{ADC}{analog-to-digital converter}
\newacronym{de}{DE}{drain efficiency}
\newacronym{pae}{PAE}{power-added efficiency}
\newacronym{sar}{SAR}{specific absorption rate}
\newacronym{mppt}{MPPT}{maximum power point tracking}
\newacronym{isi}{ISI}{intersymbol interference}
\newacronym{pps}{PPS}{pulse per second}
\newacronym{ps}{PS}{processing system}
\newacronym{pl}{PL}{programmable logic}
\newacronym{icnirp}{ICNIRP}{International Commission on Non-Ionizing Radiation Protection}
\newacronym{fd}{FD}{Front-haul distance}
\newacronym{wd}{WD}{Wireless distance}
\newacronym{ntp}{NTP}{Network Time Protocol}
\newacronym{ptp}{PTP}{Precision Time Protocol}
\newacronym{wr}{WR}{White Rabbit}
\newacronym{mm2s}{MM2S}{memory mapped to stream}
\newacronym{s2mm}{S2MM}{stream to memory mapped}
\newacronym{ber}{BER}{bit error rate}
\newacronym{mu}{MU}{multiplication unit}
\newacronym{gp}{GP}{general purpose}
\newacronym{hp}{HP}{high performance}
\newacronym{ids}{IDS}{indoor dense space}
\newacronym{dma}{DMA}{direct memory access}
\newacronym{qam}{QAM}{Quadrature amplitude modulation}
\newacronym{qpsk}{QPSK}{Quadrature  phase shift keying}
\newacronym{dc}{DC}{Divide-and-Conquer}
\newacronym{gk}{GK}{Golub-Kahan }
\newacronym{fp}{FP}{floating-point}
\newacronym{fxp}{FxP}{fixed-point}

\newcommand{\rvline}{\hspace*{-\arraycolsep}\vline\hspace*{-\arraycolsep}}

\title{Highly Parallel Singular Value Decomposition for Low-Latency MIMO Processing}

 \author{
 \IEEEauthorblockN{Sijia Cheng, Liang Liu,
 Ove Edfors, and Juan Vidal Alegría}
\IEEEauthorblockA{ Dept. of Electrical and Information Technology, Lund University, Sweden \\Email: firstname.lastname(\_lastname2)@eit.lth.se\\}

}\date{March 2025}

\begin{document}

\maketitle

\begin{abstract}
 \Gls{svd} is widely used in wireless systems, including \gls{mimo} processing and dimension reduction in \gls{dmimo}. However, the iterative nature of decomposition methods results in increased execution time as system size grows, posing challenges for real-time and low-latency applications. To address this, we analyze the latency of state-of-art \gls{svd} methods, and highlight the efficiency of a 4-step highly parallel method based on Gram matrix tridiagonalization. Furthermore, we develop a time complexity (processing latency) analysis framework with hardware profiling, allowing scalable and realistic evaluation without full implementation. The numerical results demonstrate the superior time efficiency of the selected parallel method, particularly in massive MIMO scenarios.

\end{abstract}
    \footnotetext[1]{This work is funded by the Swedish Foundation for Strategic Research (SSF) project Large Intelligent Surfaces - Architecture and Hardware, and Excellence Center at Linköping-Lund in Information Technology (ELLIIT).}
 	\glsresetall

\section{Introduction}
Computation of \gls{svd} has become a crucial task in scientific computing\cite{parallelism}.  It factorizes a matrix into orthogonal rank-1 components, each of which contributes with its own energy. This decomposition reveals the intrinsic structure of the data and can be exploited in various applications in wireless communication\cite{svdapplications}. \Gls{mimo} techniques are widely studied for their spectral efficiency and enhanced link performance, while massive \gls{mimo} can effectively exploit a large number of spatial-domain resources\cite{my}. In this context, performing \gls{svd} on the channel matrix allows diagonalizing the channel into a set of parallel, non-interfering \gls{siso} streams, which enables optimal per-stream decoding to achieve channel capacity\cite{svdmimo,mimo}.  In \gls{dmimo} applications, retaining only the most significant singular values and vectors allows reducing the data dimension at each antenna panel, so that the interconnection bandwidth to forward the data to the central processing unit can be limited with minimal information loss\cite{MA}.

Generally,  \gls{svd} is carried out by iterative operations until convergence, and the available serial algorithms have time complexity $\mathcal{O}(MK^2)$ for a matrix of size $M\times K, M \geq K$\cite{parallelism,svdmimo}.  In massive MIMO systems, the matrix dimension can scale to hundreds or even thousands, leading to high computational latency and making real-time processing challenging. Recent SVD processors are limited to small matrix sizes, and even for a $16 \times 16$ matrix decomposition, a recent published design needs 6.5k clock cycles\cite{MA}, highlighting the need for further exploration of parallelism to enable low latency processing. 


This paper first presents a highly parallel SVD method that leverages matrix size reduction and matrix symmetry to reduce processing latency. When parallelism is explored, computational complexity is not directly correlated with latency; instead, time complexity becomes the key factor. However, most existing \gls{svd} evaluation frameworks focus on computational complexity rather than time complexity\cite{framework1,frame2}, and current time complexity analysis treats all operations equally\cite{tut}, without realistic hardware models. 

To enable accurate evaluations, we develop a computational and time complexity analysis framework that considers the impact of data dependencies, processing architecture capabilities, and hardware statistics. This allows scalable and realistic algorithm evaluation without full system implementation.
By leveraging \gls{dfg} analysis, we further optimize the SVD algorithms by skipping unnecessary computations and improving data-level parallelism. The selected SVD method achieves the lowest latency among evaluated approaches, and the benefit of parallel processing becomes more pronounced in massive MIMO applications.

\section{ Parallel SVD Method}
\label{algorithms
}

\begin{figure}
    \includegraphics[width=0.8\linewidth,left]{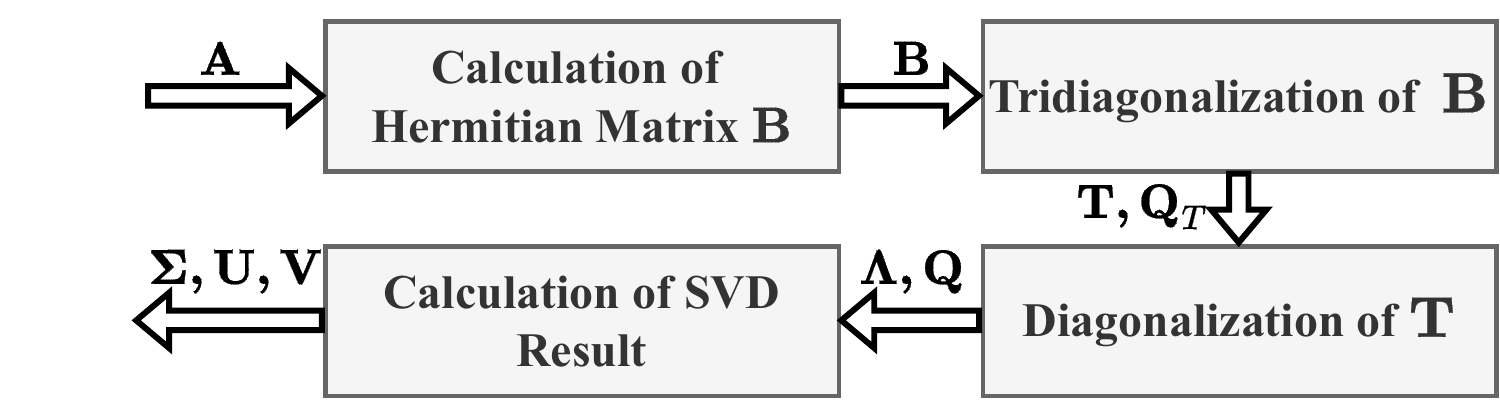}
    \caption{4-step SVD method.}
    \label{fig:4step}
    \vspace{-0.67cm}
\end{figure}
The original problem is to decompose a complex matrix $\mathbf{A}$ of size $M \times K$, where $(M\geq K$) as
\begin{equation}
    \mathbf{A} = \mathbf{U}\begin{pmatrix}
\mathbf{\Sigma}&
\mathbf{0}
\end{pmatrix}^T\mathbf{V}^H, 
    \vspace{-0.2cm}
\end{equation}
where $\mathbf{U}$ and $\mathbf{V}$ are unitary matrices of size $M\times M$ and $K\times K$, respectively, and $\mathbf{\Sigma}$ is the $K \times K$ diagonal matrix of singular values.  There are two main state-of-the-art methods
for computing the SVD. The Jacobi algorithm uses iterative rotations to successively zero out off-diagonal elements \cite{MA}, and the \gls{gk} algorithm reduces the matrix to bidiagonal form, followed by iterative diagonalization using Givens rotations\cite{book}. Both algorithms require many iterations to converge, especially for large matrices \cite{book}. 
To achieve lower processing latency, we consider the use of a 4-step SVD method, shown in Fig.~\ref{fig:4step}. It features very high data-level parallelism and allows extensive matrix dimension reduction for massive MIMO channel matrices. 
The first step converts the original $M\times K$ complex matrix $\mathbf{A}$ into a $K \times K$ Hermitian matrix $\mathbf{B}=\mathbf{A}^H \mathbf{A}$, which can be further expressed as 
$
    \mathbf{B}  
    = \mathbf{V \Sigma}^2 \mathbf{V}^H.
    $
The SVD of $\mathbf{A}$ is directly related to the \gls{evd} of  $
 \mathbf{B} = \mathbf{Q}\mathbf{\Lambda}\mathbf{Q}^H 
$, where $\mathbf{Q}$ is a unitary matrix containing the eigenvectors and $\mathbf{\Lambda}$ is the diagonal matrix of non-negative eigenvalues. Matrices $\mathbf{V,\Sigma,U}$ can then be computed as
\begin{equation}
       \mathbf{V} = \mathbf{Q},\quad
    \mathbf{\Sigma} = \sqrt{\mathbf{\Lambda}},\quad
    \mathbf{U} =[\mathbf{AV}\mathbf{\Sigma}^{-1},\mathbf{N_A}], 
        \vspace{-0.15cm}
\end{equation}
where $\mathbf{N_A}$ contains $M-K$ columns forming an orthonormal basis for the null-space of $\mathbf{A}$. 
Note that $\mathbf{N_A}$ is trivially found as the orthogonal complement of $\mathbf{AV}\mathbf{\Sigma}^{-1}$, but not directly needed in the considered applications. 
This paper assumes $\mathbf{A}$ is full rank. 
For ill-conditioned channels, only relatively significant singular values are considered for inversion. 

Operating on matrix $\mathbf{B}$ is particularly appealing in massive \gls{mimo} applications, where the number of antennas at the \gls{bs} $M$, is generally much larger than at the user side $K$, i.e., $M\gg K$\cite{frame2}. As a result, $\mathbf{B}$ has a significantly smaller dimension compared to $\mathbf{A}$, further reducing processing complexity. Moreover, $\mathbf{B}$ corresponds to the Gram matrix widely used in inter-user interference cancellation algorithms, such as zero forcing detection, enabling data sharing across processing blocks. Although some studies find that deriving singular values from eigenvalues introduces an extra error term \cite{stability,stability2}, numerical stability remains guaranteed in the \gls{mimo} applications discussed in Section \ref{application}. 

\subsection{Tridiagonalization of the Hermitian Matrix $\mathbf{B}$}

We consider the Householder transformation for tridiagonalization, as it enables parallel and vectorized operations to zero out selected components of a vector \cite{book}. In the $k^{th}$  step, the Householder vector $\mathbf{v}_k$ is constructed as
\begin{equation}
    \mathbf{v}_k= \frac{\mathbf{x}_k + e^{j \theta}\lVert \mathbf{x}_k\rVert\mathbf{e}_1 }{\lVert{\mathbf{x}_k + e^{j \theta}\lVert \mathbf{x}_k\rVert\mathbf{e}_1 } \rVert}, 
     \vspace{-0.15cm}
\end{equation}
where $\mathbf{x}_k=\mathbf{B}_{(k+1:K,k)}$, $e^{j\theta}  = x_{k,1}/|x_{k,1} |$, and $\mathbf{e}_1$ is a unit vector. By exploiting its symmetry, $\mathbf{B}_{k+1}=\mathbf{B}_{(k+1:K,k+1:K)}$ can be updated efficiently using vector-based operations
\begin{equation}
    \mathbf{B}_{k+1}' = \mathbf{B}_{k+1}-\mathbf{v}_k\mathbf{w}_k^H-\mathbf{w}_k\mathbf{v}_k^H.
    \label{eq:b} 
    \vspace{-0.15cm}
\end{equation}
where $\mathbf{w}_k= \mathbf{p}_k-(\mathbf{p}_k^H \mathbf{v}_k)\mathbf{v}_k$ and $\mathbf{p}_k=2\mathbf{B}_{k+1}\mathbf{v}_k$ \cite{book}. As $   \mathbf{B}_{k+1}'$ is symmetric, only half of the matrix needs to be calculated, reducing the computational cost. The first element of the partial column $\mathbf{x}_k$ and its mirrored row are set to $\lVert \mathbf{x}_k \lVert$, while the remaining elements are zeroed.  The transformation matrix $\mathbf{Q}_T$ is updated as $ \mathbf{Q}_T' = \mathbf{Q}_T\mathbf{P}_k^H$ where $\mathbf{P}_k = -e^{-j\theta}(\mathbf{I}-2\mathbf{v}_k \mathbf{v}_k^H)$. Finally, $\mathbf{T} = \mathbf{Q}_T^H \mathbf{B}\mathbf{Q}_T$ is tridiagonal.

\subsection{Diagonalization of the Real-valued Tridiagonal Matrix $\mathbf{T}$}

Two common algorithms for diagonalizing the symmetric tridiagonal matrix $\mathbf{T}$ are QR iterations and the \gls{dc} method \cite{book}. To reduce the number of sequential steps, the highly parallel \gls{dc} method is considered \cite{DC}. It exploits the independence to compute $K$ eigenvalues concurrently. This is particularly advantageous for large matrices, as its computational depth only scales with  $\mathcal{O}(\log_2K)$, compared to the  $\mathcal{O}(K)$ for QR iterations.

\begin{figure}
    \centering
    \includegraphics[width=0.88\linewidth]{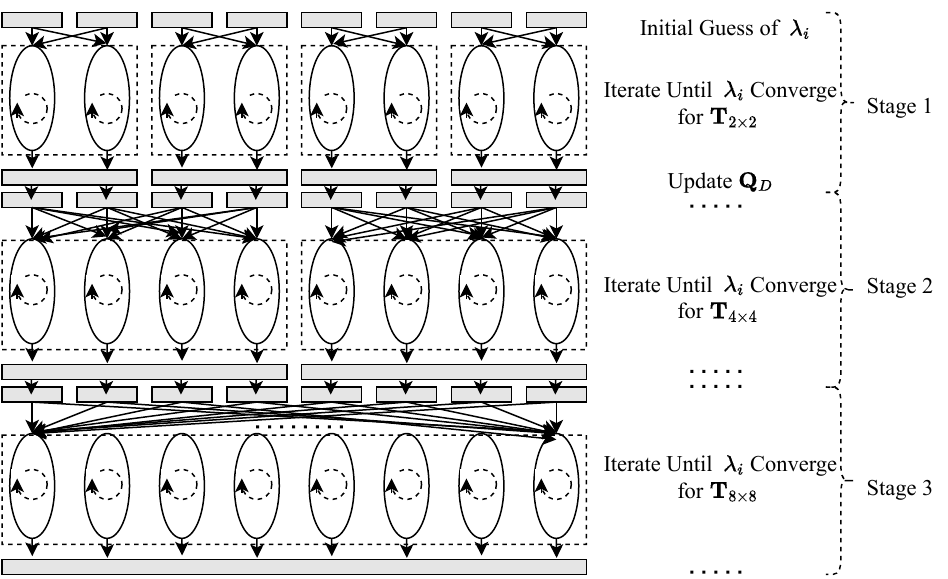}
    \caption{$8 \times 8$ \gls{dc} algorithm merge structure.}
    \label{fig:dc}
    \vspace{-0.3cm}
\end{figure}

In the divide step, 
$\mathbf{T}$ is partitioned into two smaller tridiagonal submatrices by introducing a rank-1 modification at the splitting point. Considering matrices $\mathbf{T}_1 = \mathbf{Q}_{D_1 }\mathbf{\Lambda}_1 \mathbf{Q}_{D_1 }^T$ and $\mathbf{T}_2 = \mathbf{Q}_{D_2 } \mathbf{\Lambda}_2 \mathbf{Q}_{D_2 }^T$ of dimensions $L/2 \times L/2 $, we have
\begin{equation}
 \mathbf{T}= \begin{bmatrix}
            \mathbf{T}_1&0\\0&\mathbf{T}_2
        \end{bmatrix}+ \alpha \mathbf{vv}^T =
      \mathbf{Q}_D
        (\mathbf{D}+\alpha \mathbf{uu}^T)
        \mathbf{Q}_D^T, 
         \vspace{-0.15cm}
          \end{equation}    
 where $\mathbf{v}=[0,\dots,0,1,1,0,\dots,0]^T$, 
 $\mathbf{D}=\mathrm{diag}(\mathbf{\Lambda}_1,\mathbf{\Lambda}_2)$, 
 $\mathbf{Q}_D= \mathrm{diag}(\mathbf{Q}_{D_1 },\mathbf{Q}_{D_2 })$, and $\mathbf{u} = \mathbf{Q}_D^T\mathbf{v}$. 
  This dividing process should be performed recursively until the submatrix reduces to a single element $d_i$, directly representing an eigenvalue $\mathbf{\Lambda}=[d_i]$ with corresponding eigenvector $\mathbf{e}_1$. 
 
The characteristic polynomial for $\mathbf{D}+\alpha \mathbf{uu}^T$ is derived as

\vspace{-0.3cm}\begin{equation}
 f(\lambda)= 1+\alpha \sum_{i=1}^{L} \frac{{\mu_i}^2}{d_i-\lambda} .
   \label{eq:chard}
   \vspace{-0.15cm}
\end{equation}
It can be shown that this polynomial is monotonic on $\lambda \neq d_i$ and equation $f(\lambda)=0 $ has exactly one root within each $(d_i,d_i+1)$ interval \cite{DC}. In the merge step, Newton's method is used to iteratively approximate these roots within these intervals using a second-order polynomial approximation until convergence is achieved within a given threshold. The merge structure for an $8 \times 8$ matrix is illustrated in Fig.\ref{fig:dc}. Each column in $ \mathbf{Q}_D $ is updated as $ (\mathbf{D}-\lambda_i\mathbf{I})^{-1}\mathbf{u}  $.   In the end,  $\mathbf{Q}=\mathbf{Q}_T \mathbf{Q}_D $. 

\section{Hardware Aided Framework for Processing Latency and Complexity Analysis}
\begin{figure}
    \centering
    \includegraphics[width=0.56\linewidth]{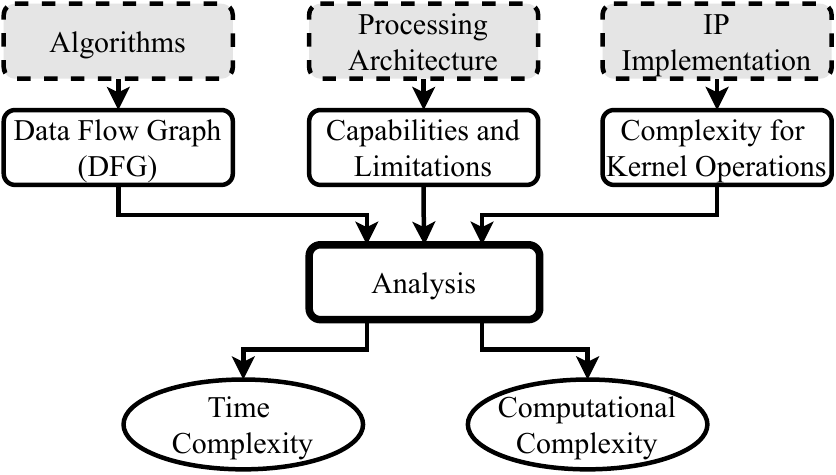}
    \caption{Time and computational complexity evaluation framework.}
\label{fig:framework}
    \vspace{-0.6 cm}
\end{figure}
To evaluate the processing latency of various SVD algorithms, we developed a hardware-supported analysis framework, illustrated in Fig.~\ref{fig:framework}. This framework consists of three aspects influencing latency at different design abstraction levels. The \gls{dfg} enables operation dependency analysis and helps identify data-level parallelism. The processing architecture description details the parallel computing capabilities of the target microarchitecture. In this paper, we assume an ideal SVD accelerator such that the analysis can focus on the algorithm features. For instance, we assume the input data is pre-buffered to avoid I/O waiting time, and the intermediate results are passed to the next-level processing units via registers to avoid memory access. Thus, the analysis focuses solely on arithmetic operation latency. Moreover, the latency of kernel arithmetic operations is provided from the actual hardware IP implementations to ensure realistic estimates.

\begin{table*}[t]
\vspace{0.1cm}
  \caption{Time and computational complexities for Householder transformation.  \label{tab:1}}
  \centering
  \begin{tabular}{|L{5.5cm}||P{1.5cm}|P{0.8cm}|P{0.8cm}|P{0.8cm}|P{2cm}|P{0.8cm}|P{0.8cm}|P{0.8cm}|}\hline
  & \multicolumn{4}{c|}{\textbf{Computational Complexity}}  & \multicolumn{4}{c|}{\textbf{Time Complexity}}\\\cline{2-9}
\textbf{Householder Transformation}& \textbf{Add} &\textbf{Mul}&\textbf{Div} &\textbf{Sqrt}& \textbf{Add} &\textbf{Mul}&\textbf{Div} &\textbf{Sqrt}\\
   \hline
$ for\quad k = 1:K-1$  & && && && &\\
1. $ \qquad  \alpha =  || \mathbf{x}_k ||$ &  $2i-1$& $2i$& &$1$&$\lceil \log_2(i-1) \rceil+2$&$1$& &$1$\\
2. $ \qquad e^{j\theta}$& &&$2$&$1$&&&&\\
 3. $ \qquad \mathbf{v}_k = \mathbf{x}_k+e^{j\theta} \alpha  \mathbf{e}_1$& $2$&$2$&&&$1$&$1$&&\\
4.  $ \qquad \mathbf{v}_k=\mathbf{v}_k/||\mathbf{v}_k||$ &$2$&$2$&$2i$&$1$&$2$&$1$&$1$&$1$\\
 5. $ \qquad  \mathbf{p}_k=2\mathbf{B}_{k+1}\mathbf{v}_k$ & $4i^2-2i$&$4i^2$ &&&$1+\lceil \log_2(i) \rceil$&$1$&&\\
 6. $ \qquad  \mathbf{w}_k=\mathbf{p}_k-\mathbf{p}_k^H\mathbf{v}_k \mathbf{v}_k^H$ & $8i-2$ &$8i$&&&$3+\lceil \log_2(i) \rceil$&$2$&&\\
 7. $ \qquad \mathbf{B}_{k+1}' =  \mathbf{B}_{k+1}-\mathbf{v}_k \mathbf{w}_k^H-\mathbf{w}_k \mathbf{v}_k^H$&$4i^2$&$4i^2$&&&$3$&$1$&& \\
8. $\qquad \mathbf{Q}'_T =\mathbf{Q}_T\mathbf{P}_k^H$&$10iK-2K$&$12iK$&&&&&&\\
 \hline
  \end{tabular}
  \begin{tablenotes}
      \small
      \item Note: $i$ is the Householder vector length, $i = K-k$. 
\end{tablenotes}

\vspace{-0.5cm}
\end{table*}

\begin{figure}
    \centering
    \includegraphics[width=0.89\linewidth]{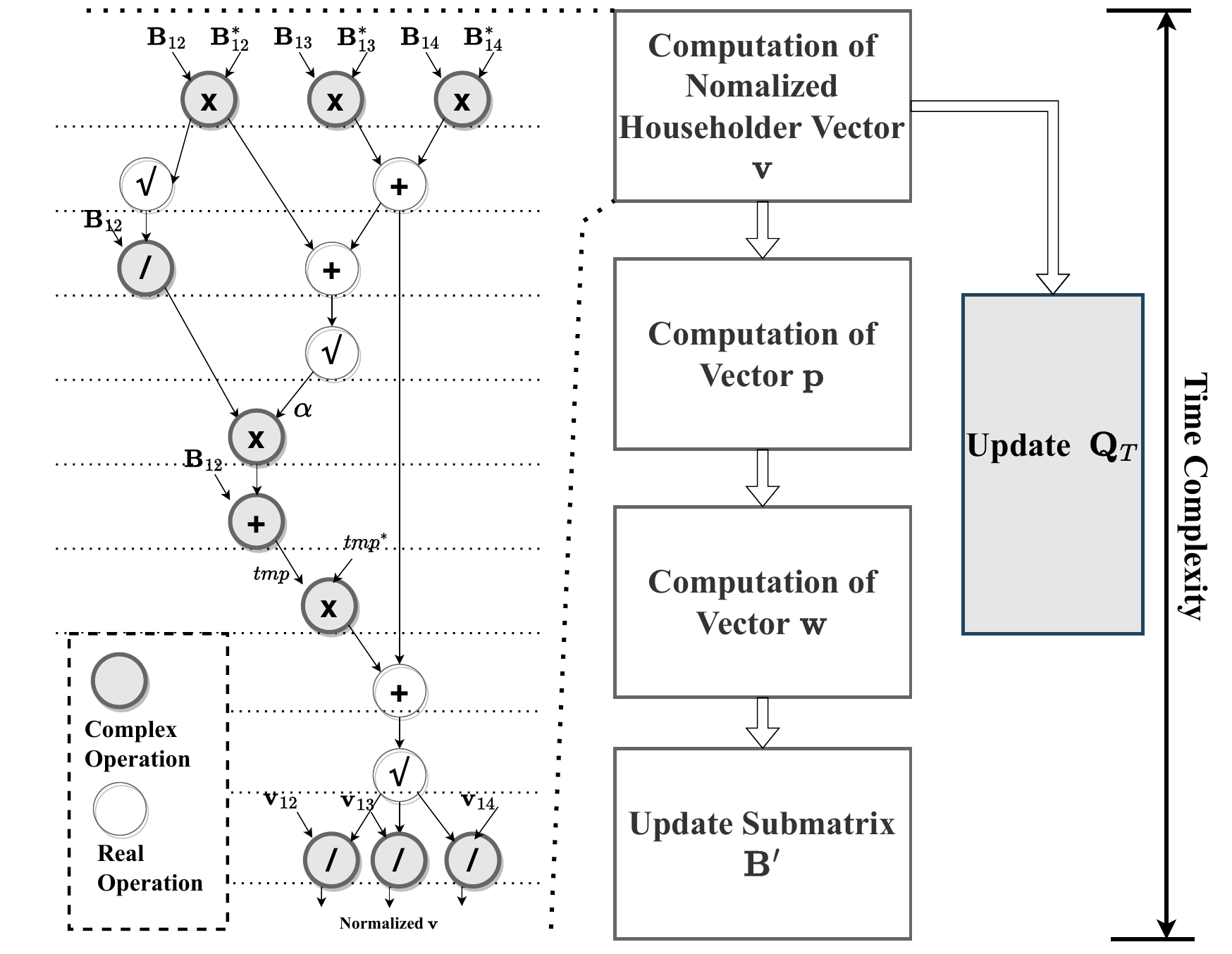}
    \caption{DFG for step 1-4 in Table \ref{tab:1}.}
    \label{fig:dfg}
    \vspace{-0.6cm}
\end{figure}

\subsection{Data Flow Analysis}
The \gls{dfg} representation of algorithms identifies parallelizable computations by leveraging its data driven nature to eliminate unnecessary waiting and to exploit data reuse. It also helps identifying the longest sequence of dependent operations, which defines the time complexity, and quantifies the number of operations to assess computational complexity.

An analysis example of Householder transformation tridiagonalization of $\mathbf{B} \in \mathbb{C}^{K \times K}$ is provided in Table \ref{tab:1}. The corresponding modularized \gls{dfg} representation for the case $K=4,k=1$ is shown in Fig.~\ref{fig:dfg}.  After DFG exploration, step 1 in Table~\ref{tab:1} first computes the sum of all elements in $\mathbf{x}_k$, excluding the first one, using an adder tree with a depth $\lceil \log2(k-1) \rceil$. The first element is then added to complete the sum. Step 2 runs in parallel with step 1 and does not affect the overall time complexity. In step 3, only the first element of $\mathbf{v}_k$ is updated, while the rest remain unchanged. As a result, in step 4, where the norm of $\mathbf{v}_k$ is calculated, the previously computed sum from step 1 can be reused to avoid redundant calculations. Optimization of steps 5-7 is based on (\ref{eq:b}). Its vector-based execution saves computation time and enhances parallel processing efficiency.

All operation counts are expressed in terms of real-valued operations. A complex multiplication is treated as four real multiplications and two real additions for computational complexity and as one real multiplication and one real addition for time complexity. When a complex number is multiplied by its conjugate, the computational complexity is reduced to two real multiplications and one real addition. Additions of multiple variants are implemented with an adder-tree for fast computation, and multiplications by powers of 2 are not counted in complexity analysis since they can be implemented using shifters.

\begin{table}[t]
\centering
\caption{Implementation result on Zynq Ultra-scale device.\label{tab:profile}}

\begin{tabular}{|cl|c|c|c|c|}

\hline
\multicolumn{2}{|l|}{}                                  & \textbf{Add}& \textbf{Mult}& \textbf{Div}& \textbf{Sqrt}\\ \hline
\multicolumn{1}{|c|}{\multirow{2}{*}{\textbf{Latency (ns)}}} & FP & $14.910$& $14.059$& $33.296$& $26.963$\\ 
\multicolumn{1}{|c|}{}                            & FxP  &$6.039$& $14.708$& $46.486$& $23.987$\\\hline
\multicolumn{1}{|c|}{\multirow{2}{*}{\textbf{\#LUT}}}      & FP & $341$& $660$& $757$& $409$\\ 
\multicolumn{1}{|c|}{}                            & FxP  & $32$& $1074$& $1242$& $352$\\\hline
\end{tabular}
\vspace{-0.5cm}
\end{table}
\subsection{Hardware Profiling of Arithmetic Operations}

We implement AMD Math Functions IPs on a Zynq Ultra-scale device to obtain a realistic evaluation of latency and complexity for different arithmetic operations. To ensure a fair comparison, all IP cores are configured as fully combinational, utilizing only \gls{lut} and avoiding \gls{dsp} blocks or other specialized resources. Table~\ref{tab:profile} shows processing latency and complexity (in terms of number of \gls{lut}s) of different arithmetic operations using both \gls{fp} and \gls{fxp} data types with 32 bits. As seen in Table \ref{tab:profile}, the data type has a significant impact on processing latency and complexity. Since \gls{fp} arithmetic offers a higher convergence rate than \gls{fxp} for SVD computation\cite{lowlatencymimo}, we use FP results for our subsequent analysis, with all the results normalized in terms of adders.
\section{Wireless Communication Use Cases\label{application}}

\begin{figure*}[t]
\centering
\begin{subfigure}[b]{0.29\linewidth}
    \includegraphics[trim={0.5cm 7.7cm 1cm 8.3cm},clip,width=\linewidth]{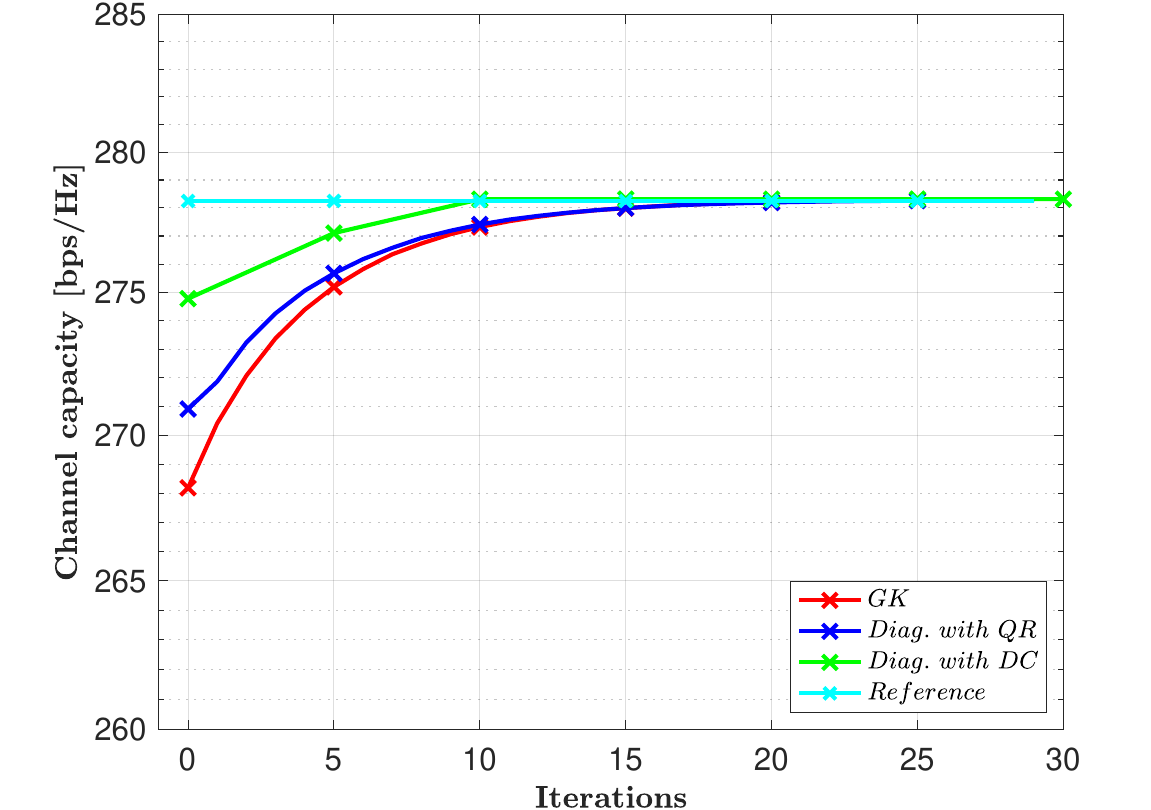}

    \caption{Channel capacity vs iterations}
    \label{fig:dmimo}
\end{subfigure}
\begin{subfigure}[b]{0.31\linewidth}
    \includegraphics[trim={0.5cm 8cm 1cm 8cm},clip,width=\linewidth]{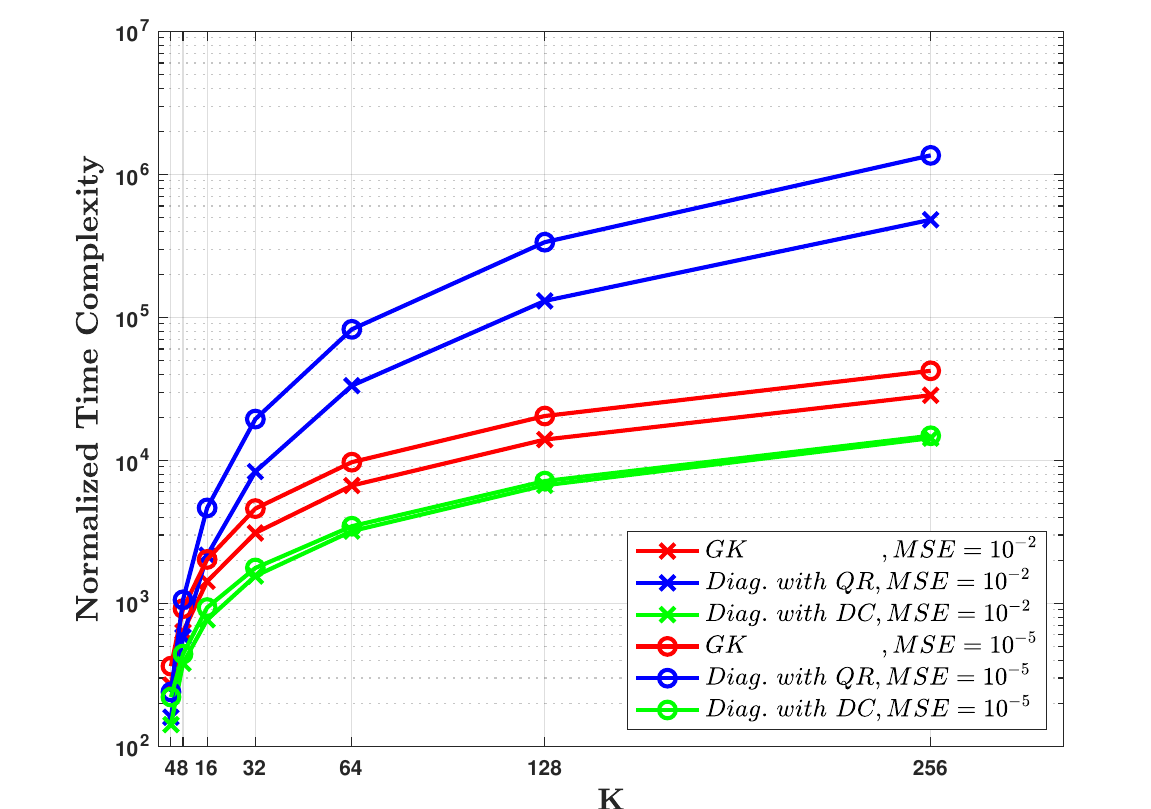}
      \caption{Time complexity comparison}
    \label{fig:timeK}
\end{subfigure}
\begin{subfigure}[b]{0.31\linewidth}
    \includegraphics[trim={0.5cm 8cm 1cm 8cm},clip,width=\linewidth]{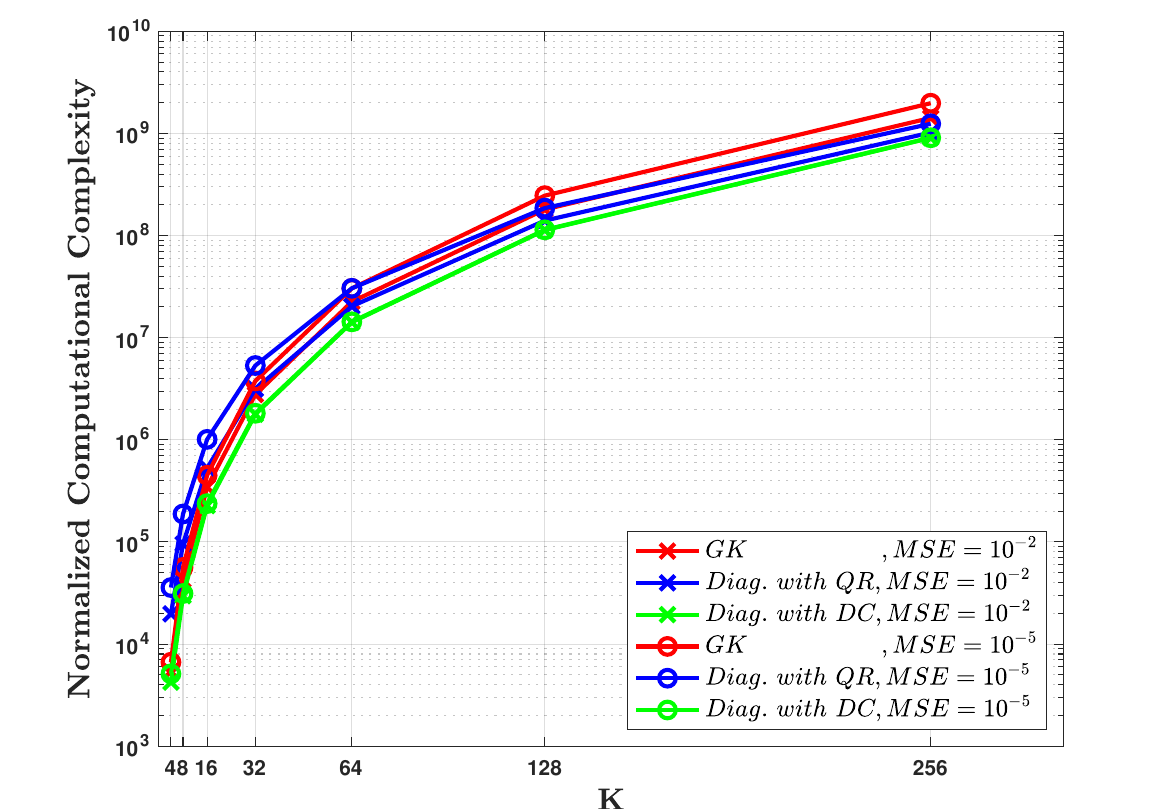}
    \caption{Computational complexity comparison}
    \label{fig:areaK}
\end{subfigure}
\caption{SVD in D-MIMO.}
\vspace{-0.5cm}
\end{figure*}
\begin{figure*}[t]
\centering
\begin{subfigure}[b]{0.29\linewidth}
    \includegraphics[trim={0.5cm 7.7cm 1cm 8.3cm},clip,width=\linewidth]{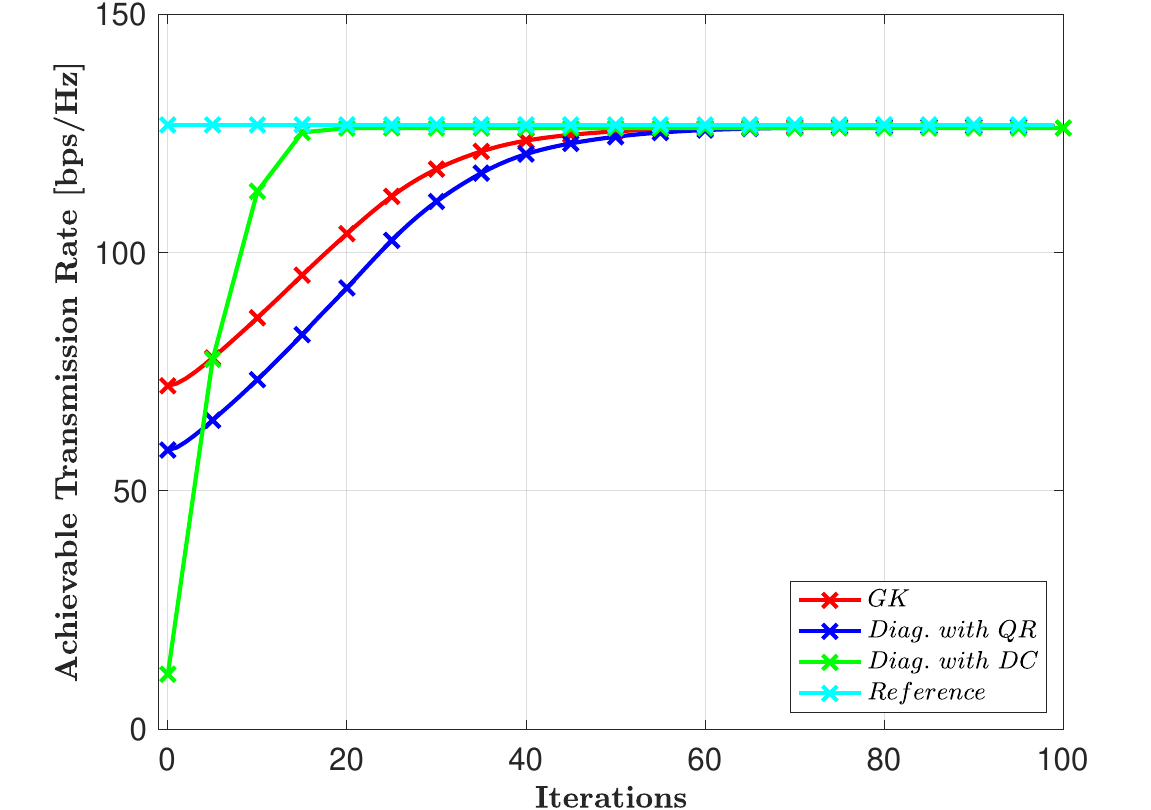}
    \caption{Achievable transmission rate vs iterations}
    \label{fig:mmimo}
\end{subfigure}
\begin{subfigure}[b]{0.31\linewidth}
    \includegraphics[trim={0.5cm 8cm 1cm 8cm},clip,width=\linewidth]{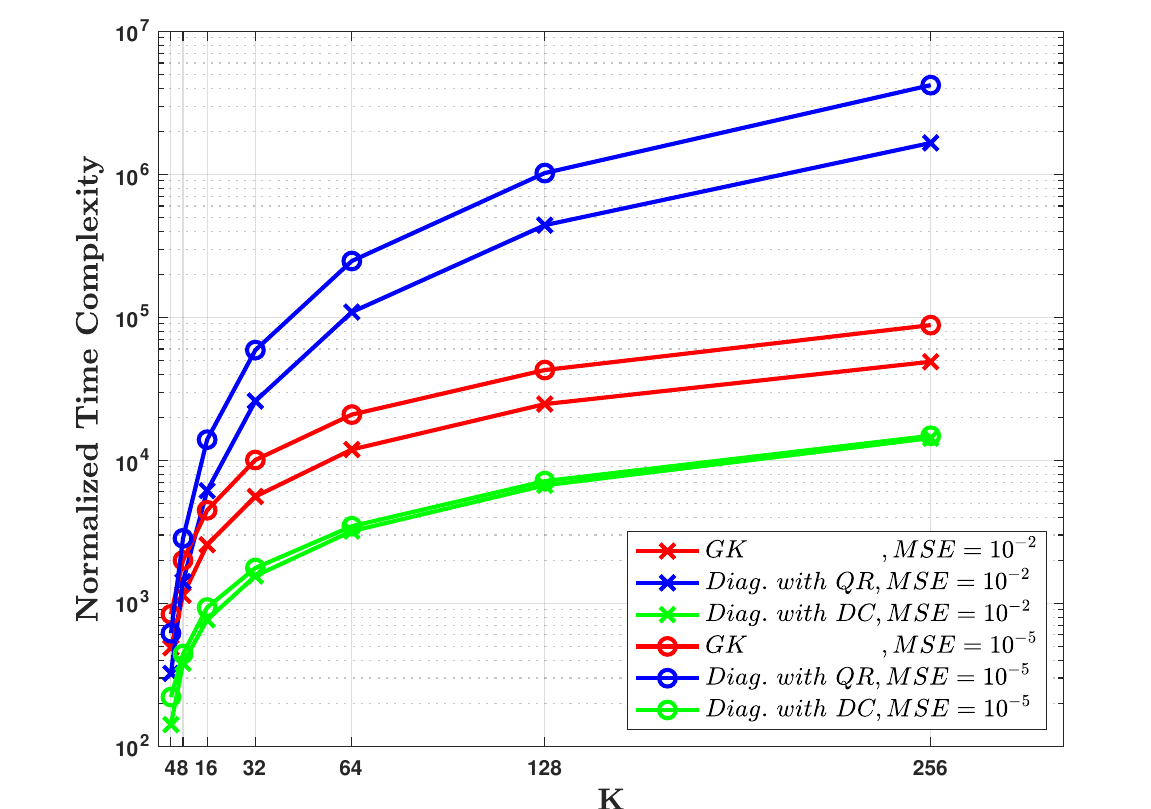}
    \caption{Time complexity comparison}
    \label{fig:time8K}
\end{subfigure}
\begin{subfigure}[b]{0.31\linewidth}
    \includegraphics[trim={0.5cm 8cm 1cm 8cm},clip,width=\linewidth]{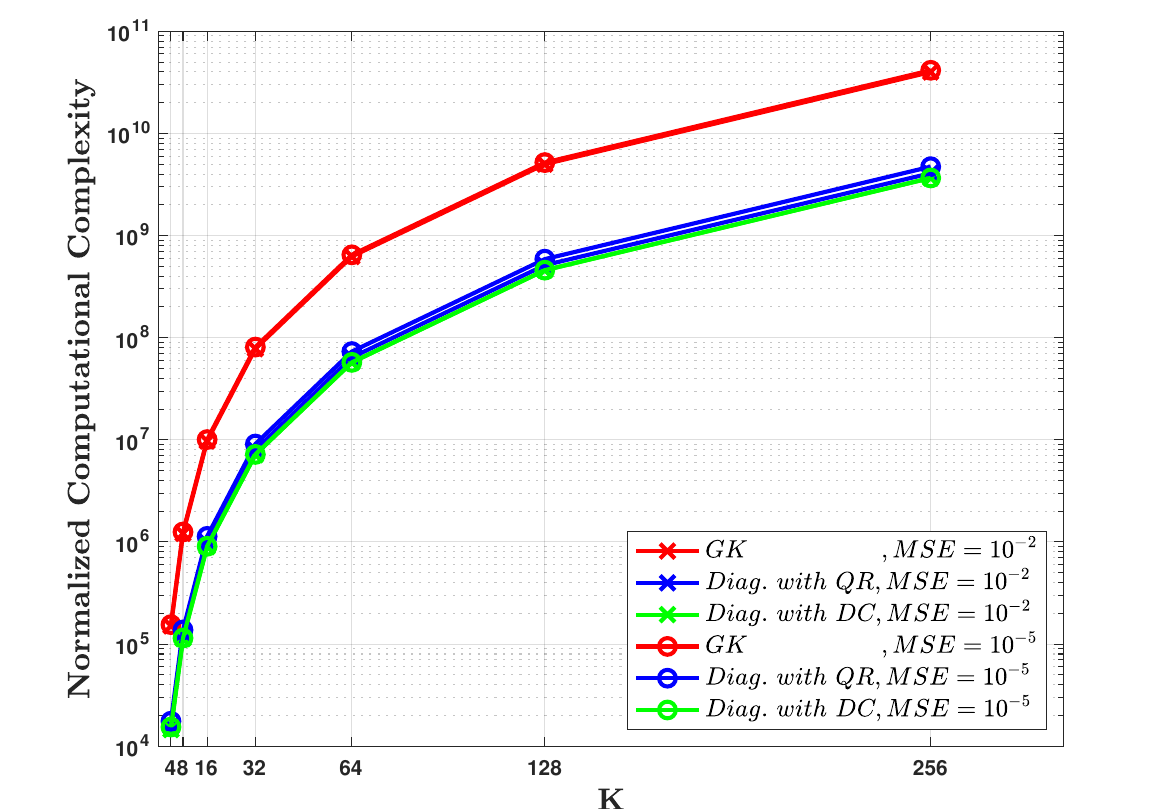}
    \caption{Computational complexity comparison}
    \label{fig:area8K}
\end{subfigure}
\caption{SVD in massive MIMO.}
\vspace{-0.6cm}
\end{figure*}


\subsection{D-MIMO}
We consider a D-MIMO system consisting of $P$ panels, each having $M$ antennas, serving $K$ single-antenna users. Assuming panel $p$ only has access to its $M\times K$ local channel $\mathbf{H}_p$, the $T\times M$ semi-unitary dimension reduction of the received vector achieving minimum information loss is given by the $T$ dominant left singular vectors of $\mathbf{H}_p$ \cite{wiffen}. Finding this dimension reduction thus requires computing the SVD of each local channel, where we assume $M\sim K$. 

The channel capacity after dimension reduction for all the mentioned SVD algorithms, including perfect-\gls{svd} as reference, is shown in Fig. \ref{fig:dmimo} for $P = 8$, $ M = 32$, $K = 32$ and $T = M/2$ for an independent and identically distributed (i.i.d) channel with \gls{snr}-per-link of $\SI{0}{dB}$. Note that, in order to achieve the channel capacity, optimal decoding should be performed under perfect knowledge of the applied transformations and of the channel while in practical settings (e.g., for individual decoding per post-processed symbol) the achievable sum rate may be degraded, as seen in the subsequent massive \gls{mimo} example. In both applications, numerical stability is maintained, as the capacity or transmission rate converges to the ideal reference, confirming that accuracy loss in selected method is acceptable. The number of iterations for DC diagonalization is computed as $iterations \times depth$, where, in this case, $depth = \log_2K =5$. Time complexity and computational complexity when $M = K$ is shown in Figs.~\ref{fig:timeK} and \ref{fig:areaK} for different \gls{mse} requirements on the singular values.

For comparison, the \gls{gk} method is evaluated, as it requires fewer computations than the Jacobi method\cite{MA}. Note that the concurrency between successive iterations can be explored in the \gls{gk} diagonalization step, as Givens rotations only affect a $2 \times 3$ submatrix \cite{GKpipelined}. We also evaluate the use of QR iterations in the diagonalization step in our proposed method, which includes a refinement that eliminates trivial multiplications with 0s and 1s. Fig. \ref{fig:timeK} shows that the proposed parallel method with DC diagonalization achieves a significantly lower time complexity, approximating $\mathcal{O}(K)$. This improvement comes from parallel eigenvalue computations at each recursion stage, greatly reducing latency. While the number of stages scales with $\mathcal{O}(\log_2K)$, the number of iterations required for accurate decomposition increases with matrix size. Meanwhile, all evaluated algorithms exhibit a computational complexity of $\mathcal{O}(K^3)$ as shown in Fig. \ref{fig:areaK}.

\subsection{Massive MIMO}

 Consider a massive MIMO \gls{bs} with $M$ antennas serving $K$ single-antenna users through a narrowband channel, with $M\gg K$, leading to a channel matrix $\mathbf{H}\in \mathbb{C}^{M\times K}$. Achieving channel capacity under individual user decoding relies in performing \gls{svd} of the channel matrix $\mathbf{H}=\mathbf{U}\mathbf{\Sigma}\mathbf{V}^H$
 \cite{mimo}. In case of having an inaccurate \gls{svd}, the achievable user rate may be computed 
from the post-processed \gls{sinr} after diagonalizing the channel with the erroneous estimates of $\mathbf{U}$ and $\mathbf{V}$ \cite{mimo}, resulting in degradation from extra interference.

Fig. \ref{fig:mmimo} shows the achievable rate with $K=16,M=128$ under an i.i.d channel with an \gls{snr}-per-link of $\SI{0}{dB}$. Time complexity and computational complexity for the case where $ M\gg K$ is shown in Fig. \ref{fig:time8K}  and Fig. \ref{fig:area8K}. Compared to the case where $M \sim K$, the advantage of the proposed parallel method with DC diagonalization becomes more obvious. In this scenario, the dimension reduction achieved by transforming the channel matrix into the Gram matrix is more beneficial. It is worth noting that using the \gls{dc} method for diagonalization outperforms the QR iterations because its complexity does not increase significantly with increased $M$. For QR iterations, the number of iterations required to achieve a certain accuracy increases due to the presence of larger super-diagonal and sub-diagonal elements in the tridiagonal matrix $\mathbf{T}$, slowing down convergence. In contrast, the \gls{dc} method does not suffer from this drawback, as it efficiently finds eigenvalues through recursive approximation rather than iterative rotations.

\section{Conclusion and Future Work}
This paper presents a highly parallel \gls{svd} method and evaluates its time and computational complexity using a novel latency analysis framework, with a special focus on MIMO applications. Compared with other commonly used algorithms, the selected parallel method achieves lower latency by leveraging data-level parallelism and optimizing data dependencies. Future work includes hardware implementation to validate performance and explore further architectural improvements.

\bibliographystyle{IEEEtran}
\bibliography{references}

\end{document}